\documentclass[11pt]{article}
\usepackage[top=0.9in, left=1in, bottom=0.9in]{geometry}
\usepackage{hyperref}
\usepackage{upgreek}
\pdfoutput=1

\begin{document}
\title{Cellular Blood Flow}
\author{Jonathan R.~Clausen, Daniel A.~Reasor Jr., and Cyrus K.~Aidun \\
\\\vspace{6pt} George W.~Woodruff School of Mechanical Engineering, \\
Georgia Institute of Technology, Atlanta, GA, USA}

\maketitle

%% The abstract (in this file, and that submitted as text to arXiv) should include the exact phrase "fluid dynamics video" or "fluid dynamics videos"
\begin{abstract}
The fluid dynamics video that is presented here outlines recent advances in the simulation of multiphase cellular blood flow through the direct numerical simulations of deformable red blood cells (RBCs) demonstrated through several numerical experiments. Videos show particle deformation, shear stress on the particle surface, and the formation of particle clusters in both Hagen--Poiseuille and shear flow.
\end{abstract}

% main text
\subsection*{Introduction}

This video includes demonstrations of dilute and dense suspensions of RBCs in wall-bounded shear and pressure-gradient driven pipe flow. The video, available in hi-resolution and web-friendly formats, can be found at 	\href{http://ecommons.library.cornell.edu/handle/1813/14097}{http://ecommons.library.cornell.edu/handle/1813/14097}

The animations presented in the movie submission are results from the direct numerical simulation of deformable suspensions using a hybrid lattice-Boltzmann/finite element method that is coupled with a particle--particle and particle--boundary contact model~\cite{macmeccan2009sdp}. The code has also been optimized to run on as many as 65,536 cores of the IBM \emph{Blue Gene/P} architecture~\cite{clausen2009ppl}. In order to simulate high volume fraction (hematocrit) simulations of deformable particles in wall-bounded or cylindrical domains, a seeding method is used. In this procedure, the particles are initialized at 30\% of their final size and grown to their full size rigidly, allowing particle--particle and particle--wall interactions using the lubrication and contact model previously referenced. Each RBC is composed of 504 linear triangular shell elements (254 nodes) resulting in a 762$\times$762 finite element matrix for each particle updated with Newmark's time integration method. Physically, RBCs are biconcave in shape with a maximum diameter of $8\,\upmu$m enclosed by an elastic membrane with an effective elastic shear modulus of $5.7\times10^{-3}\,\mbox{dyn cm}^{-1}$~\cite{waugh1979trb} and a bending stiffness of $2.2\times10^{- 
12}\,\mbox{dyn cm}$~\cite{hwang1997edl}, which surrounds a liquid hemoglobin with a viscosity of 6~cP. The RBCs are suspended in blood plasma with a viscosity of 1.2~cP. The relative nondimensional elastic parameter is the elastic capillary number, $\mbox{Ca}_G \equiv \mu \dot{\gamma} a/G_m$ where $\mu$ is the fluid viscosity, $\dot{\gamma}$ is the local shear rate, $a$ a length scale based on the particle size, and $G_m$ is the elastic shear modulus.

\subsection*{Hagen--Poiseuille of Cellular Blood Flow}
	In these simulations, the capillary number of the particles increases with the radial distance from the axis of the tube, reaching a maximum at the tube wall. The pressure-gradient for these simulations is constant and is mimicked by a body force in the axial direction. Periodic boundary conditions are implemented at the tube inlet and outlet to allow the flow of RBCs exiting the domain to re-enter. The fluid domain for these simulations is 128$\times$128$\times$512 (8.39 million grid points). Clusters of RBCs form in relatively high volume fraction Hagen--Poiseuille flow, and these clusters increase the relative suspension viscosity. Statistics captured include the number of clusters, average cluster size, and the effect of hematocrit and capillary number. The formation of three of these clusters is visualized at different time steps for the simulation described here. In Hagen--Poiseuille flow, the shear stress is at a maximum at the tube wall. As a result, the RBCs nearest the tube wall experience the highest rate of shear. This relatively high shear results in higher capillary numbers for the RBCs near the tube wall. The particles also tend to migrate away from areas of high shear and form a core in the center of the tube where the RBCs remain relatively undeformed. 
%
%	Through the study of four different hematocrit values, the ensemble average of the radial hematocrit distribution reaches a maximum at the core and the rate of decline is different for each of these values. At a constant pressure-gradient, an increase in the hematocrit value, causes a decrease in the flow rate through the tube by increasing the effective viscosity of the bulk fluid. This is not only due to the presence of the suspensions themselves, but also due to the hemaglobin fluid within each RBC that has a viscosity of 6cP which is five times larger than the plasma viscosity of 1.2cP. 

\subsection*{Linear Shear Flow of Red Blood Cells and Capsules}
	Both wall-bounded and unbounded shear simulations are presented in this movie. The RBC simulation is in a cubical fluid domain of 256$\times$256$\times$256 (16.8 million grid points) with 2470 deformable RBCs (40\% hematocrit) at $\mbox{Ca}_G = 0.02$. The movie portrays the different deformed shapes of the biconcave RBC geometries~\cite{goldsmith1979fbe}. The unbounded shear simulation is of deformable spherical capsules ($\mbox{Ca}_G$ = 0.02) with RBC membrane properties in a 144$\times$144$\times$144 domain. The video shows cluster formation along the compressional axis and cluster separation along the extensional axis with corresponding changes in stress on the particle's surface.
	
\subsection*{Acknowledgements}
J.R. is funded by IPST @ GT, and D.R. is funded by the U.S. Department of Defense through the SMART fellowship program. These simulations were performed through the use of TeraGrid resources at Purdue made available by the National Science Foundation and IBM \emph{Blue Gene/P} system at the Argonne Leadership Computing Facility. 
\bibliographystyle{unsrt}		% BIB Style

\end{document}